\documentclass[12pt,preprint]{aastex}
\defcitealias{ng04}{Paper I}

\begin{document}
\title{Fitting Pulsar Wind Tori. II. Error Analysis and Applications}
\author{C.-Y. Ng\altaffilmark{1} \& Roger W. Romani\altaffilmark{2}}
\altaffiltext{1}{School of Physics, University of Sydney, NSW 2006, Australia}
\altaffiltext{2}{Department of Physics, Stanford University, Stanford, CA 94305}
\email{ncy@physics.usyd.edu.au, rwr@astro.stanford.edu}

\begin{abstract}
We have applied the torus fitting procedure described in Ng \& Romani (2004)
to PWNe observations in the \emph{Chandra} data archive.
This study provides quantitative measurement of the PWN geometry and
we characterize the uncertainties in the fits, with statistical errors
coming from the fit uncertainties and systematic errors estimated by
varying the assumed fitting model. The symmetry axis $\Psi$ of the
PWN are generally well determined, and highly model-independent. We often
derive a robust value for the spin inclination $\zeta$. We briefly discuss
the utility of these results in comparison with new radio and high energy pulse
measurements.
\end{abstract}


\section{Introduction} 
One of the greatest success of the Chandra X-ray Observatory (CXO) is the
discovery of equatorial tori and polar jet structures in many pulsar wind
nebula (PWN) systems. It is now believe that these features are common
among young neutron stars. In the \citet{ree74} picture, when the highly
relativistic pulsar wind decelerates in the external medium, a termination
shock is formed at a characteristic scale
\[r_t =\left(\frac{\dot E}{4\pi c \eta P_{\rm ext}} \right )^{1/2} \ , \]
where $\dot E$ is the pulsar spin-down power and $\eta$ is the filling factor.
In general, if the pulsar is subsonic in the ambient medium, i.e.
$ P_{\rm ext} \geq P_{\rm ram}= 6\times 10^{-10}nv^2_7\;\mathrm
{g\;cm^{-1}s^{-2}}$ for a pulsar speed $10^7v_7\;\mathrm{cm\;s^{-1}}$ through
a density of $n\,m_p\;\mathrm{cm^{-3}}$, a toroidal shock structure is expected;
faster objects produce bow shock nebulae. Indeed, many young pulsars still
inside their high pressure supernova remnant birth sites do show such
toroidal symmetry. The best-known example is the PWN around the 
\object{Crab pulsar} as observed by the \emph{CXO} \citep{wei01}.
Recently, several relativistic MHD models, e.g.\ \citet{kom03} and 
\citet{del06}, have shown how such toroidal structure can form if the 
pulsar wind has a latitudinal variation.

\citet{ng04} (hereafter \citetalias{ng04}) developed a fitting procedure to measure the
3D orientation of the pulsar wind torus and applied to a few X-ray observations.
While this simple geometrical model does not capture the fine details 
of the MHD simulations, is does allow one to extract the torus (and hence pulsar spin)
orientation from relatively low signal-to-noise ratio (S/N) data. 
\citetalias{ng04} also gave quantitative estimates for the statistical errors
arising from Poisson statistics. However the systematic errors due to
unmodeled components such as jets or background were neglected. For bright objects e.g.\
the Crab and Vela pulsars, the S/N is high and such systematic errors dominate.
In this study, we attempt to quantify these systematic errors and apply the fitting 
to more \emph{CXO} PWN observations, thus providing a more comprehensive study.  

\section{Pulsar Wind Torus Fitting}
\subsection{Brief Review}

In \citetalias{ng04}, we developed a generic torus fitting procedure to capture the
characteristic structure of PWNe. Here we give a brief summary
of the model: consider a torus in 3D with radius $r$ and Gaussian cross
section of thickness $\delta$.  The emissivity is described by
\[I_0\propto \exp\left[-\frac{{z^\prime}^2+(\sqrt{{x^\prime}^2+{y^\prime}^2}-r)^2}{2\delta^2}\right] \]
where $z^\prime$ is the direction along the torus axis and
the observer line of slight lies in the $y^\prime z^\prime$-plane.  
As the post-shock flow in the PWNe is expected to be mildly relativistic,
the torus intensity in the observer's frame is Doppler boosted by
\[ I \propto (1-\mathbf{n}\cdot \mbox{\boldmath$\beta$})^{-(1+\Gamma)} I_0 \ . \]
Here we assume a constant flow speed {\boldmath$\beta$}={\boldmath$v$}$/c$
radially from the pulsar
in the equatorial plane. Of course this does not capture the
complex latitudinal structure and speed variations seen in the 
MHD simulations \citep[e.g.][]{del06}, but it does give a characteristic
speed and captures the gross Doppler boosting.
During the fitting we leave {\boldmath$\beta$} as a free parameter, but assume
a fixed photon spectral index $\Gamma=1.5$ in the rest frame, which is a typical
value for PWNe. We have also tried other values of $\Gamma$, e.g.\ $\Gamma=1.2$
and $\Gamma=1.9$ as reported for the Vela and Crab pulsars respectively 
\citep{kar04,mor04}. The geometrical parameters are quite insensitive to $\Gamma$.
The phenomenological speed {\boldmath$\beta$} is slightly affected (with 
$\sim \delta \mbox{\boldmath$\beta$} \approx -0.15\,\delta\Gamma$),
so large spectral index variations could induce trends in the fit speed.
Altogether this gives
\[ I(x^\prime,y^\prime,z^\prime)= \frac{N}{(2\pi\delta)^2r}
 \left(1 - \frac{y^\prime \sin \zeta}{\sqrt{{x^\prime}^2
+{y^\prime}^2}}|\mbox{\boldmath$\beta$}|\right)^{-(1+\Gamma)}
\exp\left[-\frac{{z^\prime}^2+(\sqrt{{x^\prime}^2+{y^\prime}^2}-r)^2}{2\delta^2}\right] \ ,\]
where $N$ is the total number of counts in the torus and
$\zeta$ is the inclination of the torus.

In order to compare with data, the 3-D torus is projected onto the sky plane.
We set up the coordinate system for the image plane with $y$ along celestial
north, $z$ along the observer line of sight and the CCD in the $xy$-plane. Then the
transformation between the frames is given by
\[ \left ( \begin{array}{c} x^\prime \\ y^\prime\\ z^\prime
\end{array} \right ) =
\left ( \begin{array}{ccc}
-\cos \Psi & -\sin \Psi & 0 \\
\sin \Psi \cos \zeta & -\cos \Psi\cos \zeta & \sin \zeta \\
-\sin \Psi \sin \zeta & \cos \Psi \sin \zeta & \cos \zeta
\end{array} \right) 
\left ( \begin{array}{c} x \\ y\\ z \end{array} \right ) \ , \]
where $\Psi$ is the position angle (N through E) of the projected torus axis.
The counts in each CCD pixel (x,y) result from the integration of emissivity
through the line of sight
\[ C(x,y)=\sum_z I(x^\prime,y^\prime,z^\prime) \ . \]

The basic torus model is characterized by the parameters $\Psi$, $\zeta$,
$r$, $\delta$ and {\boldmath$\beta$}. In addition, a point source and constant background
are also included in the fit. The counts in each component are determined by
the fit, while the total counts in the model is constrained to match the
image counts. In a few cases where twin tori are observed, there is an extra
fitting parameter $d$ for the separation. For simplicity we assume
a symmetrical offset along the torus axis, with the pulsar in the middle.
A few other cases have inner and outer tori, at different $r$.

\begin{figure}[!ht]
\epsscale{0.5}
\plotone{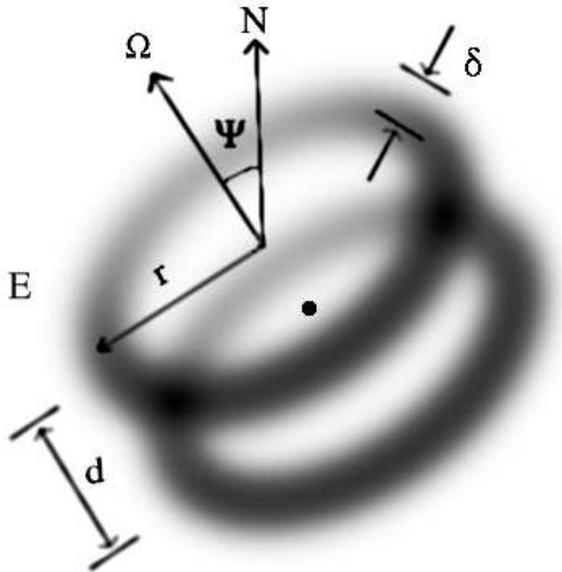}
\caption{\label{fig1}Illustration of a twin torus, labeled with the fitting parameters.}
\end{figure}

To estimate the goodness of fit for particular model parameters, a 
likelihood function is defined using
Poisson statistics. The probability of observing $d_{xy}$ counts out of the
expected $C_{xy}$ counts in pixel $(x,y)$ is given by
\[ P(d_{xy})=\frac{C^{d_{xy}}_{xy}e^{-C_{xy}}}{d_{xy}!} \ .\]
In the high-count limit, we switch to Gaussian statistics for
$C_{xy} \geq 20$, hence the likelihood function passes to the usual
$\chi^2$ statistics. The best-fit parameters are then determined by
the maximum likelihood method, i.e. minimization of the Figure-of-merit function,
which is defined as the negative log of the likelihood function summed over all
image pixels.

Once the best-fit parameters are obtained, statistical errors in the fit are
estimated by Monte Carlo simulations. Poisson realizations of the best-fit model
are simulated and fitted, and the confidence interval of each parameter is
obtained from their distributions. To estimate the systematic errors, we blank out
different regions in the data (i.e.\ omit those pixels from the fit), then
re-fit the model by varying only one parameter at a time while fixing the others at
their best-fit values. The changes in the best-fit parameter values thus provide
estimates of the uncertainties contributed by un-modeled features in the data.
Note that in addition to the single parameter systemetic errors computed here,
covariance between the parameters could also lead to larger errors
(e.g. the covariance of {\boldmath$\beta$} with $\Gamma$).

\subsection{Implementation}

In this study we went through the \emph{CXO} data archive and identified a list of
observations with apparent toroidal termination shocks in PWNe (Table~\ref{tab1}).
There are a few other objects with jets and/or possible tori (e.g.\ PSR B1509$-$58),
but for these cases closer examinations suggest that the pulsar motions maybe
trans- or super-sonic with respect to the surrounding medium,
hence static torus fits are not appropriate.
The data are cleaned up with standard reduction procedures using CIAO and the background
light curves are examined to filter out periods that suffered from strong particle
flares. Then we remove the pixel randomization for the ACIS observations and apply
an algorithm by \citet{mor01}, which corrects the positions of the split pixel events,
to improve the spatial resolution. Finally we select the appropriate energy range and
binning to optimize the S/N of the PWNe, separating out the soft background from the
SNR and field stars. To model the point source, a high S/N PSF model is simulated with
the Chandra Ray Tracer (ChaRT) at the detector source position, using the 
source spectrum reported in literature.

\begin{deluxetable}{cccccc}
\tablecaption{\label{tab1}Archival dataset used in this study.}
\tablewidth{0pt}
\tabletypesize{\scriptsize}
\tablehead{
\colhead{Pulsar} & \colhead{SNR} & \colhead{ObsID} &
\colhead{Instrument} & \colhead{Exposure (ks)}}
\startdata
J0205+6449 & 3C 58 & \dataset[ADS/Sa.CXO#obs/3832]{3832},
\dataset[ADS/Sa.CXO#obs/4382]{4382}, \dataset[ADS/Sa.CXO#obs/4383]{4383} & ACIS & 350 \\
J0537$-$6910 & N157B & \dataset[ADS/Sa.CXO#obs/2783]{2783} & ACIS & 50 \\
B0540$-$69 & SNR 0540$-$69.3 & \dataset[ADS/Sa.CXO#obs/132]{132},
\dataset[ADS/Sa.CXO#obs/1735]{1735}, \dataset[ADS/Sa.CXO#obs/1736]{1736},
\dataset[ADS/Sa.CXO#obs/1727]{1727}, \dataset[ADS/Sa.CXO#obs/1738]{1738},
\dataset[ADS/Sa.CXO#obs/1741]{1741}, \dataset[ADS/Sa.CXO#obs/1745]{1745} & HRC & 75 \\
J1124$-$5916 & G292.0+1.8 & \dataset[ADS/Sa.CXO#obs/1953]{1953} & HRC & 50 \\
B1800$-$21 & \nodata & \dataset[ADS/Sa.CXO#obs/5590]{5590} & ACIS & 30 \\
J1833$-$1034 & G21.5$-$0.9 & \dataset[ADS/Sa.CXO#obs/1230]{1230},
\dataset[ADS/Sa.CXO#obs/1554]{1554},\dataset[ADS/Sa.CXO#obs/159]{159},
\dataset[ADS/Sa.CXO#obs/2873]{2873},\dataset[ADS/Sa.CXO#obs/3700]{3700} & ACIS & 56
\enddata
\end{deluxetable}

\begin{figure}[!ht]
\epsscale{0.78}
\plotone{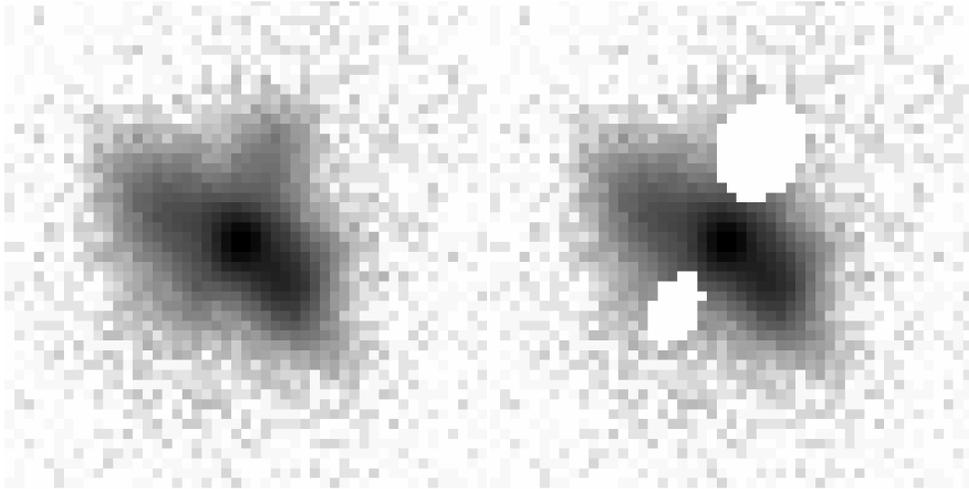}
\caption{\label{fig2}\emph{Left: Chandra} ACIS
image of PSR B0540$-$69; \emph{right}: same image with the jet regions excised.}
\end{figure}

In the fitting procedure, a simulated annealing package \citep{pre92} is
employed for multidimensional minimization of the Figure-of-merit function, with the
initial parameters chosen by eye. After the best-fit model is obtained, we simulate
500 Monte Carlo Poisson realizations and estimate the statistical errors from the
parameter distributions accordingly. The $1\sigma$ confidence interval, which
corresponds to 68\% confidence level, are reported. To estimate the systematic errors,
we begin by blanking out the jet regions in the observations,
since they are the most obvious un-modeled  features and are present in many systems.
As an example, the image of PSR B0540$-$69 with jet regions removed is shown in
Figure~\ref{fig2}. The golden section search in one dimension \citep{pre92} is then
employed for re-fitting the model. This procedure gives systematic error estimates for
the Crab, \object{Vela}, PSRs \object{J2229+6113}, \object{B1706$-$44},
\object{J2021+3651} and \object{B0540$-$69}. However for PSRs \object{J1930+1852}
and \object{J0205+6449}, the jets are relatively faint and well-separated from the PWNe.
Therefore, removing the jets alone does not change the parameter values significantly.
We believe this underestimates the systematic errors and contributions from other
un-modeled features should also be considered. Therefore, we obtained another estimate
by removing the point source in the data. The results reported for these two objects
are the combination of the two systematic error estimates, added in quadrature.

\subsection{Individual Objects}

In the following we discuss the fits of some individual objects.

\emph{PSR B0540$-$69:} The combined HRC exposure shows polar jet structure in the PWN.
In addition, a bright blob SW of the point source is also observed.
We have tried blanking out the blob in the fit, and found that the result is 
insensitive.

\emph{PSR J0205+6449:} Similar to the Crab pulsar, a double tori structure is
observed in this PWN. However, in this case the outer torus is very diffuse
and without a clear boundary. In order to have a better constraint in the
position angle, we fit the two tori simultaneously with the same value of $\Psi$.

\emph{\object{PSR J1124$-$5916:}}
While the PWN image in the initial {\it CXO} HRC observation showed noticeable
ellipticity \citep{hug03}, the elongation was not clear in a subsequent ACIS exposure.
This could possibly due to the large off-axis angle degrading the PSF.
Therefore in this study we fitted the HRC image only. A scheduled deep ACIS
observation could help to resolve this structure.

\emph{\object{PSR J0537$-$6910:}}
A cometary trail is observed in this PWN system \citep{che06},
which is modeled in the fit
by a rectangular background region with constant brightness. Comparison with
the best-fit parameters without the background provides an estimate for the
systematic errors.

\emph{\object{PSR B1800$-$21:}}
Although the 30\,ks ACIS exposure only detects a total of $\sim$200 counts (0.8-6\,keV)
for the point source and the PWN, the latter clearly shows extended structure with 
a well-defined symmetry axis. Interestingly, the direction of elongation is in
general orthogonal to the proper motion claimed in \citet{bri06}. Hence,
employing the torus fitting to obtain a quantitative measurement of the $\Psi$
accuracy is valuable.

\emph{\object{PSR J1833$-$1034:}}
As it's host SNR is one of the calibration objects for \emph{CXO}, the pulsar has
been observed repeatedly. We combined all the on-axis and in-focus ACIS
observations in the data archive and obtain a total exposure of 56\,ks.

\emph{PSR J0538$-$2817:}
In a short 20\,ks ACIS exposure, diffuse emission around the pulsar was fit to
a simple equatorial flow. However, while a deeper 100\,ks observation \citep{nge07}
showed extended emission along a similar axis, it was not unambiguously possible
to decide whether this was an edge-on equatorial torus or a polar jet.  To be 
conservative, we do not include this object in the present analysis.

\section{Results and Discussions}

\begin{figure}[!ht]
\epsscale{1.0}

\plottwo{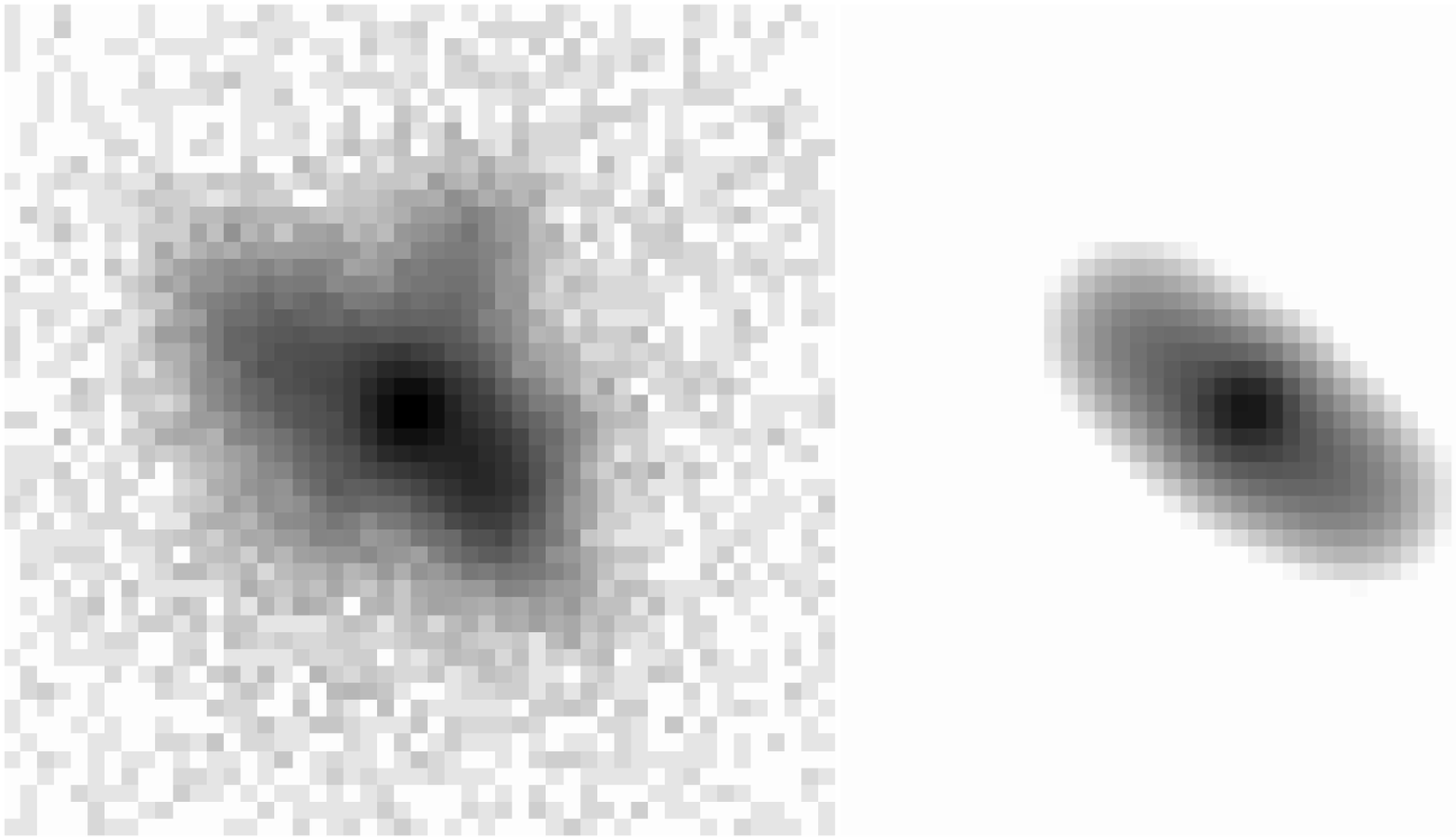}{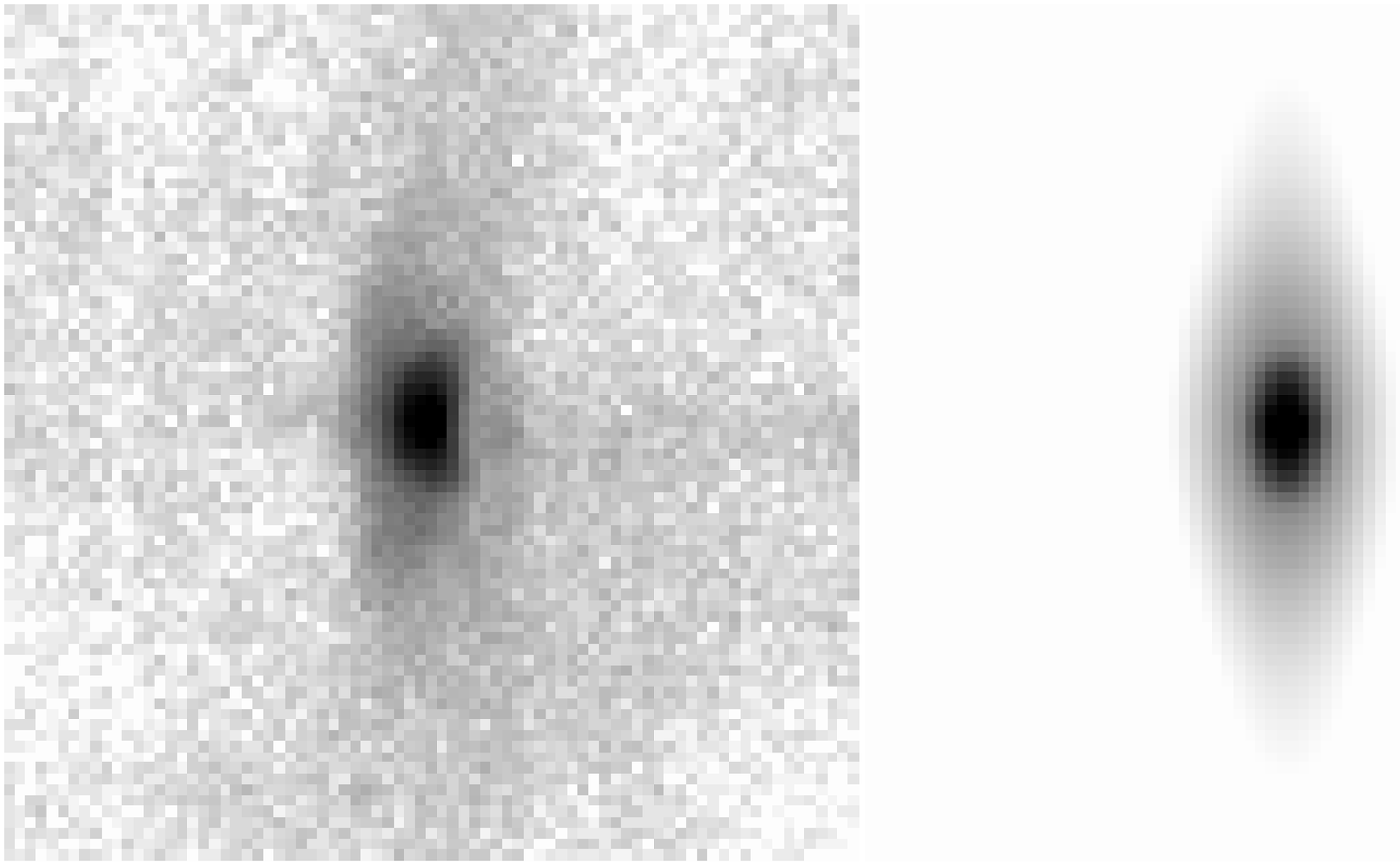}

\plottwo{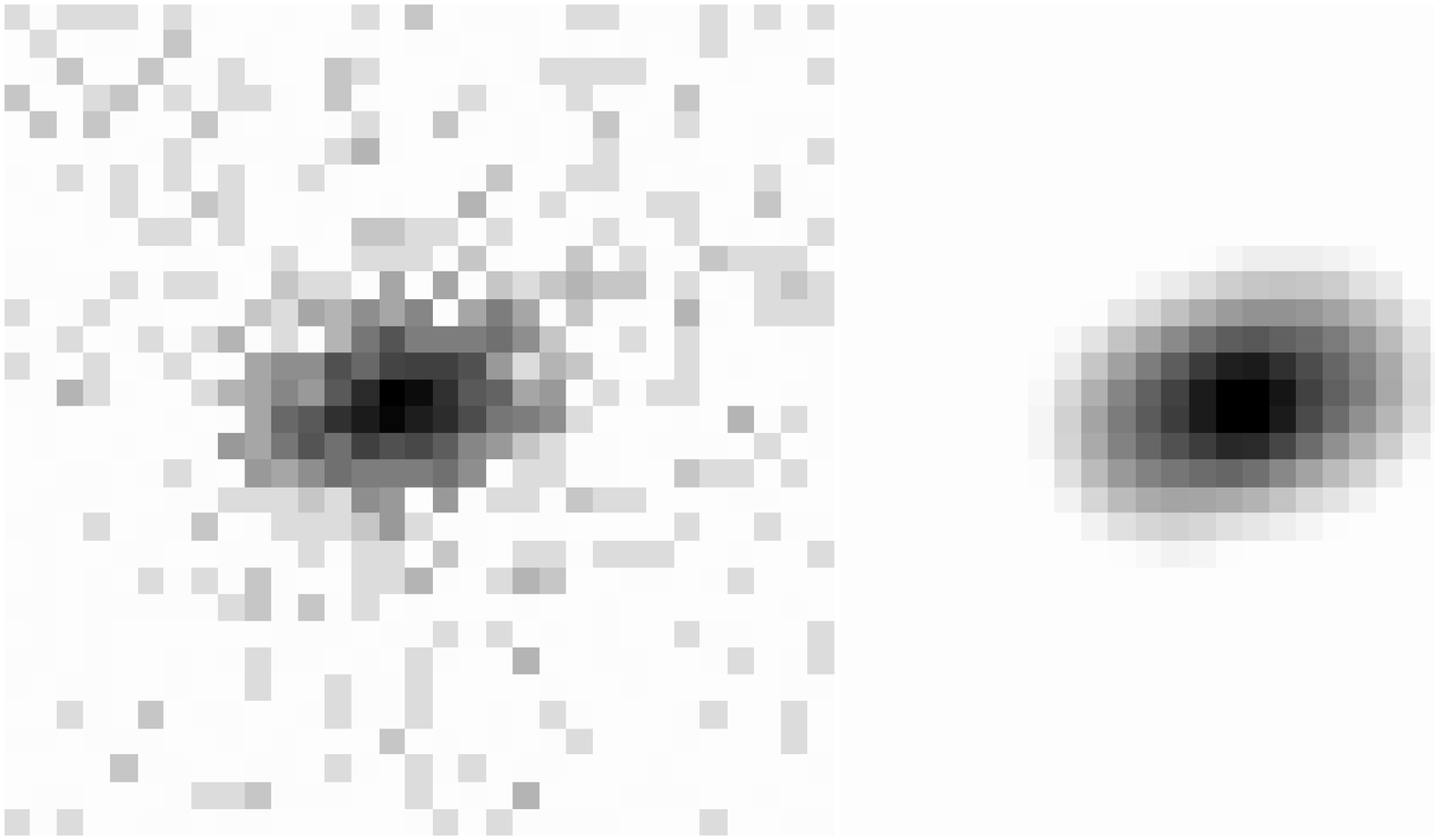}{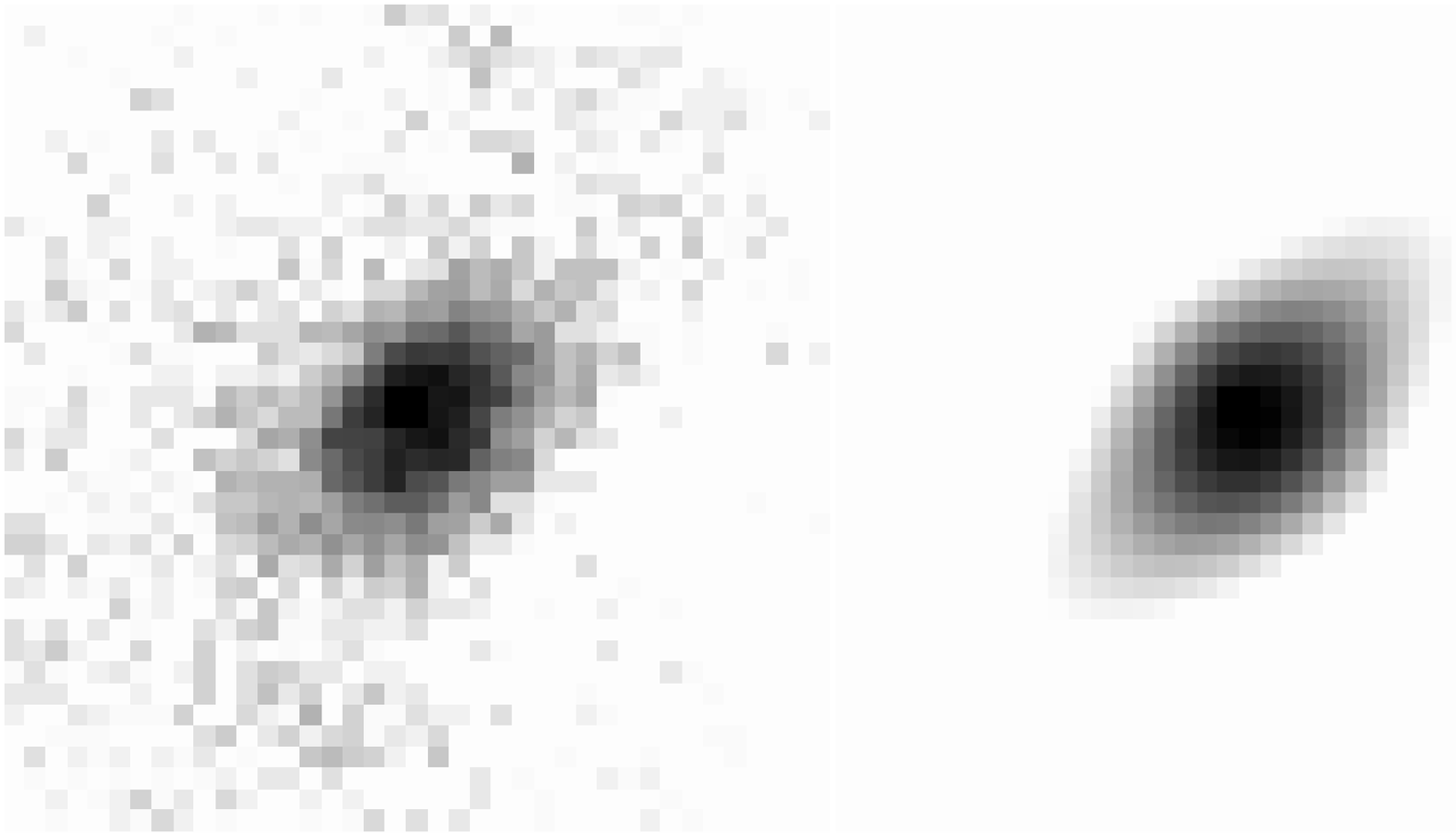}

\plottwo{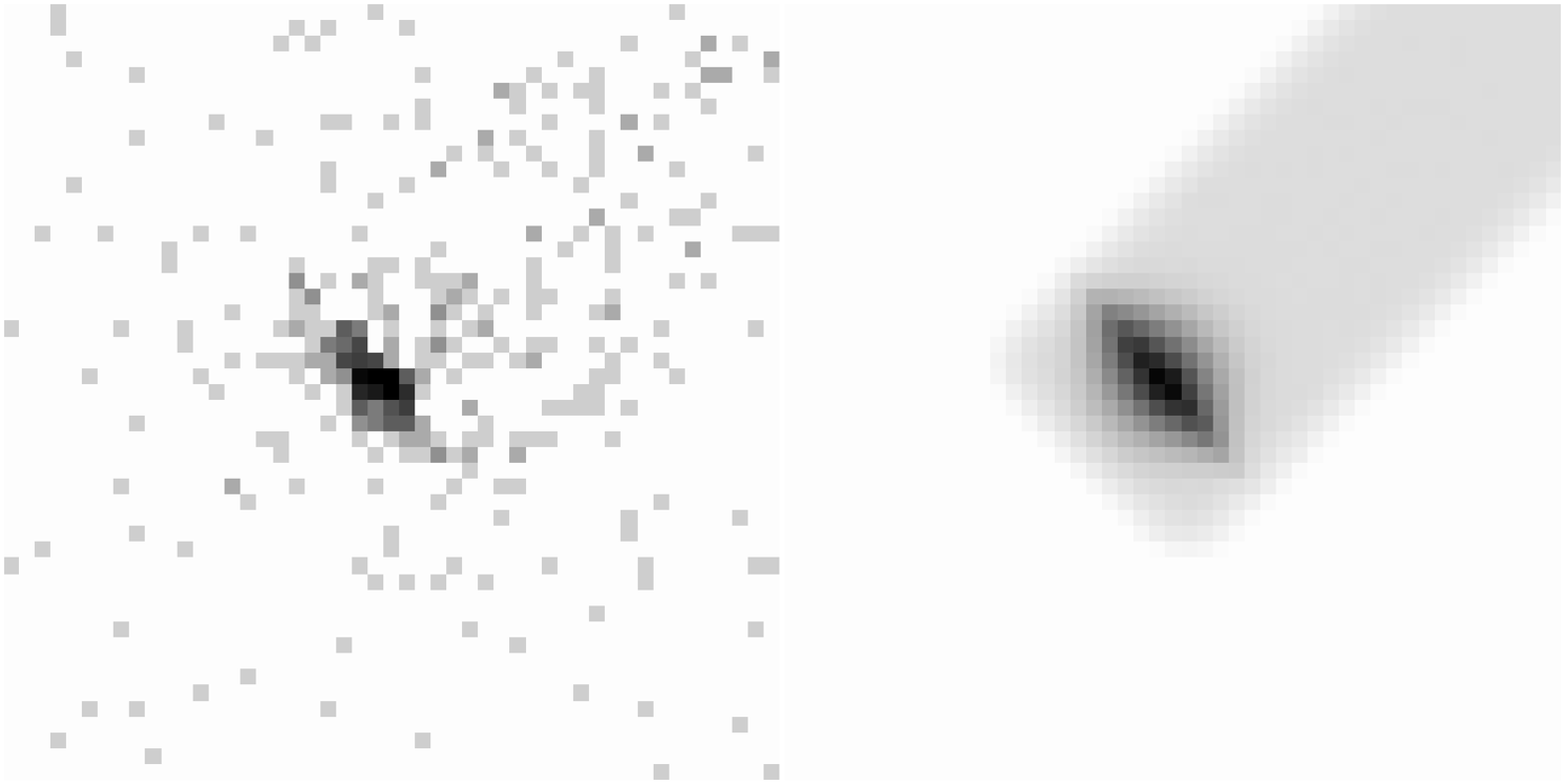}{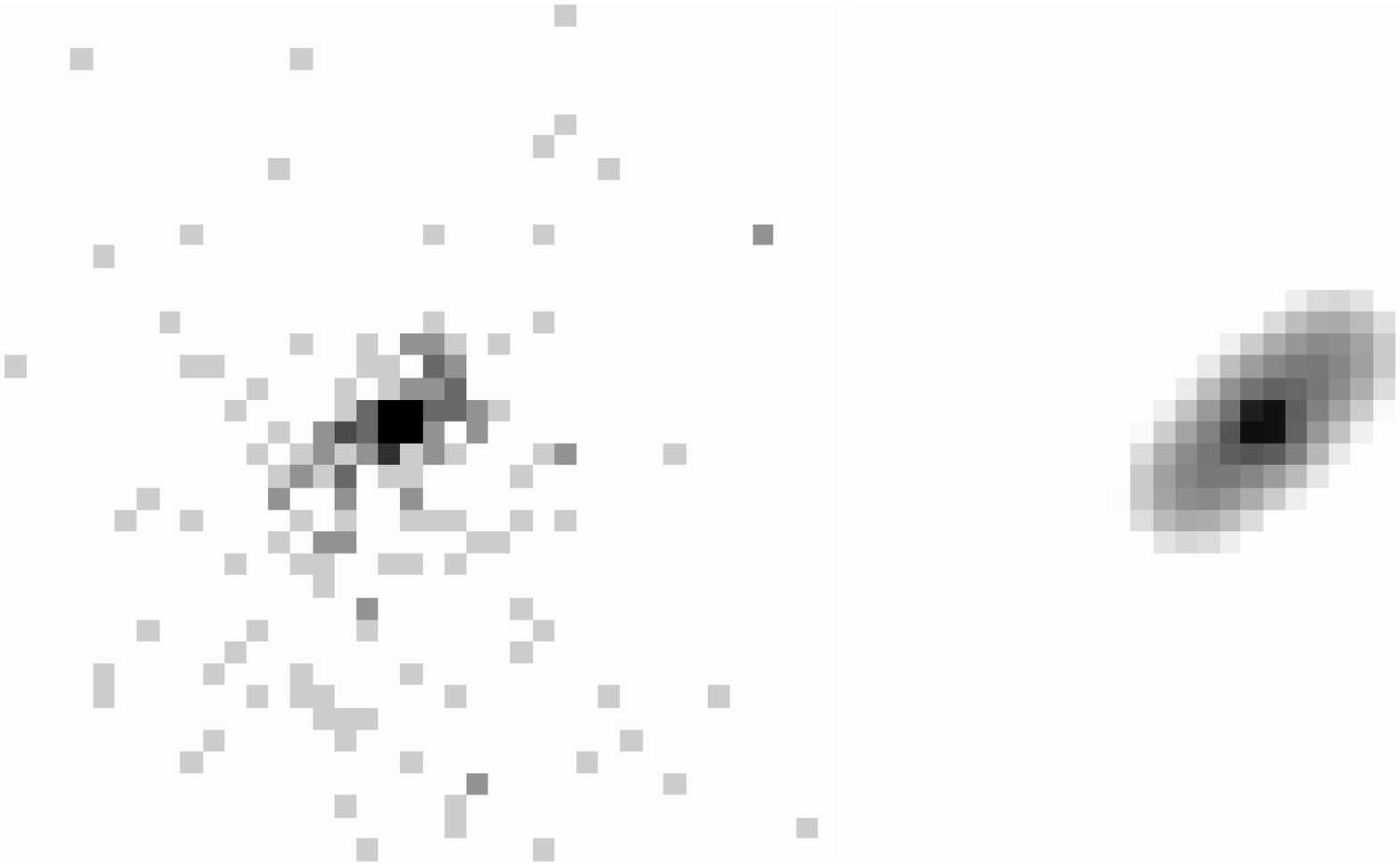}

\caption{\label{fig3}New fitting results in this study.
\emph{Chandra} data (\emph{left}) compared with the best-fit models (\emph{right}) for PSRs 
B0540$-$69, J0205+6449, J1124$-$5916, J1833$-$1034, J0537$-$6910 and B1800$-$21(\emph{left to right,
top to bottom}).}
\end{figure}

Table~\ref{tab2} lists the best-fit results with corresponding statistical and systematic
errors. We report the systematic errors for $\Psi, \zeta, r$ and the separation only, since
they are the most interesting parameters. Systematic errors in other parameters have similar
fractional values when compared to the best-fit parameters. Note that for the faint objects,
including PSRs J1124$-$5916, B1800$-$21 and J1833$-$1034, the detailed structures are
not resolved and the large statistical errors dominate over the systematic errors.
Therefore only the former are reported here. The best-fit models for the new
objects in this study are shown in Figure~\ref{fig3}, the rest may be found in
\citetalias{ng04}, Figure 2 and the references listed in Table~\ref{tab2}.

\begin{deluxetable}{lcccccccc}
\tablecaption{\label{tab2}Best-fit `torus' parameters with 1$\sigma$ uncertainties. The
statistical and systematic errors are listed as the first and second error terms
respectively. We report the systematic errors for selected objects in $\Psi, \zeta$, $r$ and {\boldmath$\beta$} only.}
\tablewidth{0pt}
\rotate
\tabletypesize{\scriptsize}
\tablehead{
\colhead{object} & \colhead{$\Psi (\arcdeg)$} & \colhead{$\zeta (\arcdeg)$} & \colhead{$r (\arcsec)$} & \colhead{$\delta (\arcsec)$} & \colhead{\boldmath$\beta$} & \colhead{sep ($\arcsec$)} & \colhead{PS/torus (Cts)} & \colhead{Ref.}
}
\startdata
Crab (inner) & $124\pm0.1\pm0.1$ & $61.3\pm0.1\pm1.1$ & $15.6\pm0.03\pm0.1$ & 1.5* & $0.49^{+0.005}_{-0.006}\pm0.006$ & \nodata & \nodata /$1.0\times 10^5$ & 1 \\
Crab (outer) & $126.31\pm0.03\pm0.11$ & $63.03^{+0.02}_{-0.03}\pm1.3$ & $41.33^{+0.02}_{-0.03}\pm0.2$ & 5.9* & $0.55\pm0.001\pm0.0004$ & \nodata & \nodata /$1.1\times 10^7$ & 1 \\
Vela & $130.63^{+0.05}_{-0.07}\pm0.05$ & $63.6^{+0.07}_{-0.05}\pm0.6$ & $21.25^{+0.03}_{-0.02}\pm0.5$ & 3.0* & $0.44^{+0.004}_{-0.003}\pm0.008$ & $11.61\pm0.03\pm0.4$ & \nodata /$1.3\times 10^6$ & 1 \\
J1930+1852 & $91^{+4}_{-5}\pm1.1$ & $147\pm3\pm3$ & $4.6\pm0.1\pm0.14$ & $1.1\pm0.1$ & $0.62^{+0.04}_{-0.03}\pm0.1$ & \nodata & 1701/602 & 1 \\
J2229+6114 & $103\pm2\pm1.6$ & $46\pm2\pm6$ & $9.3\pm0.2\pm0.3$ & 2.5* & $0.49\pm0.02\pm0.09$ & \nodata & 2221/1113 & 1 \\
B1706$-$44 & $163.6\pm0.7\pm1.6$ & $53.3^{+1.6}_{-1.4}\pm2.9$ & $3.3^{+0.08}_{-0.06}\pm0.1\pm0.01$ & 1.0* & $0.70\pm0.01$ & \nodata & 2897/1221 & 2 \\
J2021+3651 & $45\pm1.3\pm0.6$ & $79\pm1\pm2$ & $8.0\pm0.2\pm0.4$ & 1.2* & $0.64\pm0.02\pm0.02$ & $3.7^{+0.2}_{-0.1}\pm0.07$ & 235/751 & 3 \\
J0205+6449 (inner) & $90.3^\dagger\pm0.2\pm0.4$ & $91.6\pm0.2\pm2.5$ & $2.08^{+0.04}_{-0.02}\pm0.01$ & 0.74* & $0.66^{+0.01}_{-0.01}\pm0.009$ & \nodata & 5080/39730 & 4 \\
J0205+6449 (outer) & $90.3^\dagger\pm0.2\pm0.4$ & $90.56^{+0.07}_{-0.05}\pm0.01$ & $16.3^{+0.3}_{-0.1}\pm0.1$ & 2.0* & $0.88^{+0.004}_{-0.001}\pm0.0007$ & \nodata & 5080/29503 & 4 \\
J0537$-$6910 & $131\pm2\pm0.9$ & $92.8^{+0.7}_{-0.8}\pm0.5$ & $4.0^{+0.5}_{-0.2}\pm0.3$ & $0.47\pm0.04$ & $0.86\pm0.02\pm0.08$ & \nodata & 102/179 & 4 \\
B0540$-$69 & $144.1\pm0.2\pm0.8$ & $92.9\pm0.1\pm0.6$ & $2.35\pm0.02\pm0.06$ & 0.66* & $0.64\pm0.01\pm0.02$ & \nodata & 13442/50215 & 4 \\
J1124$-$5916 & $16\pm3$ & $105\pm7$ & $0.9\pm0.1$ & $0.5^{+0.02}_{-0.05}$ & $0.15^{+0.06}_{-0.04}$ &
   \nodata & 390/552 & 4 \\
B1800$-$21 & $44\pm4$ & $90\pm2$ & $3.1^{+0.4}_{-0.5}$ & 0.74* & $0.69^{+0.05}_{-0.12}$ & \nodata & 18/88 & 4 \\
J1833$-$1034 & $45\pm1$ & $85.4^{+0.2}_{-0.3}$ & $5.7\pm0.2$ & $\sim1.2$ & $0.86\pm0.01$ & \nodata & 304/6541 & 4
\enddata
\tablenotetext{*}{ -- held fixed in the fit.}
\tablenotetext{\dag}{ -- fitted simultaneously.}
\tablerefs{Best-fit parameters and statistical errors are from: (1) \citet{ng04},
(2) \citet{roe05}, (3) \citet{hes04}, (4) -- this work. \\All systematic
errors are new in this work.}
\end{deluxetable}

As shown in the table, the fits for Crab and Vela are dominated by
systematic errors and the statistical errors are negligible.
For the other objects, the two are comparable in $\Psi$ and $r$, but the
systematic errors for $\zeta$ are in general much larger. This suggests the fit is most
robust in obtaining $\Psi$ and $r$, as they are largely determined by the shape of the torus. 
On the other hand, $\zeta$ depends sensitively on the relative brightness of both sides of
the torus; thus, removing the jet structure can change the results significantly.

\subsection{Pulsar Jets}
As mentioned in the previous section, polar jets are observed in many of the pulsars
in this study. In general, the jets' position angle is in reasonable agreement with
$\Psi$ axis determined from the torus geometry. Since these are also relativistic,
relative Doppler boosting could in principle provide a second estimate of spin 
inclination $\zeta$ and flow speed {\boldmath$\beta$}. However, two effects complicate this. 
First, {\boldmath$\beta$}
might be quite different in the jets as suggested in the MHD simulations \citep{del06}.
Second, the jets are in several cases distinctly curved, and repeated exposures of
Crab and Vela indicate that this bending in the plane of the sky changes with time. 
Since we expect similar curvature along the line-of-sight, and even quite
small changes in inclination strongly affect the brightness ratios, interpretation of 
the jet boosting is more difficult than that of the tori.

\subsection{Pulsar Kick Connection}
One direct application of the torus fitting is measurement of the neutron star's spin
axis orientation. The alignment angle between the pulsar spin and velocity vectors
provides insights into the supernova core collapse physics. For a simple picture of
a kick acting at a fixed spot on the neutron star surface, a longer kick duration
gives a better alignment as a result of rotational averaging. \citet{ng07}
performed simulations of neutrino kick models and extracted constraints on the physics 
parameters including the kick direction, magnitude and timescale by comparing 
torus fitting angles with proper motion estimates and other pulsar data.
As the sample of X-ray PWNe with modeled torus structure increases, the statistical 
constraints on such kick models should rapidly improve.

\subsection {Radio Polarization} 
There are a few objects in our list which also have independent estimates for 
the spin geometry from radio polarimetry. \citet{joh05}, in particular, have
recently made considerable progress in measuring $\Psi$ by determining the polarization
at the phase of closest approach to the magnetic pole and correcting for Faraday 
rotation to obtain the absolute position angle on the sky. Unfortunately, radio pulsar emission
is complex and can switch between orthogonal modes; this implies a $\pi/2$ ambiguity 
in the spin axis orientation. Nevertheless, careful analysis can give useful position
angles. At present only three of our pulsars have a $\Psi$ from radio measurements.
\citet{joh05} observed the radio polarization for the Vela pulsar and obtain a P.A.\ of
$126\fdg8\pm1\arcdeg$.  For PSR B1706$-$44, the result is controversial: while 
\citet{wan06} report a P.A.\ of $162\fdg0\pm10\fdg0$, \citet{joh05} do not give a value,
since in new observations the phase of closest approach is difficult to locate.
Finally, \citet{kot06} measure the radio polarization of the PWN in PSR
J2229+6114 and suggest a symmetry axis at P.A.$\sim93\arcdeg$. While these three 
values are in reasonable agreement with the $\Psi$ from the X-ray symmetry axis,
more data are needed for a serious comparison. 

Unfortunately, interpretation of
the polarimetry appears to be particularly challenging for the young pulsars showing
X-ray tori (this may be related to their powerful accelerators in relatively small
magnetospheres). For example, the main radio pulses in the Crab appear at the same phases
as the high energy emission and may indicate high-altitude emission. A classic 
polarization sweep may only be found for the `precursor', a low frequency component.
However, measurement of radio polarization for more objects would be very useful.
Comparison with the X-ray P.A.s can reveal orthogonal emission in the radio, e.g.
\citet{hgh01}, and may help to calibrate the radio polarization techniques, which may then
be applied to older pulsars. PSRs J2020$+$3651 and B1800$-$21 seem amenable to such
analysis.

	In principle, for a simple dipole geometry it is also possible to measure the
magnetic pole impact parameter $\beta$ (from the polarization PA sweep rate at closest
approach) and the magnetic inclination $\alpha$ (from the overall shape of the polarization
position angle sweep), when the polarization is well measured over a large range of
pulsar phase. In practice, while $\beta$ can be fairly well determined, the $\alpha$
estimates are often poor and subject to large systematic errors. The Rotating Vector Model
(RVM) for such fits was developed for Vela, so not surprisingly there are angles
for this pulsar. Table 3 gives the values from \citet{joh05}. However, the sum
$\alpha+\beta=\zeta=43^\circ-6\fdg5 \sim 37^\circ$ is in poor agreement with the
X-ray inclination of $64^\circ$. It is difficult to interpret the {\it CXO} arc structure seen
for Vela with such a small inclination. Perhaps the best approach is to adopt $\zeta$
from the PWN structure, when available, and $\beta$ from radio polarization. For Vela
this would imply $\alpha \sim 70^\circ$.

\subsection {The $\gamma$-ray Connection}

	Radio (coherent emission) pulse profiles are often quite complex, and may 
depend on the dense plasma and complex field structure at their relatively low 
emission altitudes.  High energy (incoherent emission) profiles should be simpler, 
especially if they originate at high altitude
where only dipole field components are expected to persist. \citet{rom95}
found that for the generic class of outer magnetosphere models, the
$\gamma$-ray pulse is usually a double peak, with a peak separation $\Delta$ and 
phase lag $\delta$ from the radio pulse primarily determined by $\zeta$, although
the actual values do depend on $\alpha$ and the gap width, as well. Other models,
employing emission from both hemispheres \citep[e.g.][]{dr03} can have more
complex light curves with different phase relationships. However, there are some 
generic predictions of these models: $\gamma$-ray peaks are strong and wide when 
$\zeta \sim \pi/2$, they converge as $\zeta$ approaches $\sim \pi/4$, and are weak
or absent for small angles. 

In Table \ref{tab3}, we list the observed GeV pulse widths ($\Delta_\gamma$)
for Crab, Vela and PSR B1706$-$44 and tentative widths for two other pulsars
discovered after the end of the EGRET mission. The final column ($\Delta_{OG}$)
lists approximate
high energy pulse widths (or ranges) from the contours in \citet{rom95}, figure 4.
In most cases, a range is allowed -- independent (e.g. radio) values for $\beta$
would make width predictions. Note that for Crab and Vela, this model tends to
predict somewhat narrower pulses than those seen. Also the tentative detection of
PSR J2229+6114 has a wider pulse than predicted for its small $\zeta$.
These larger angles may be easier to accommodate in two-pole models.
At present there is sufficient latitude in the model predictions to allow a wide
range of pulse shapes. However, new $\gamma$-ray observations with AGILE and, 
especially, GLAST should provide many high quality pulse profiles. We expect to
have X-ray measurements of $\zeta$ for many of these young pulsars. If radio observations
can supply $\beta$, the geometry is determined and comparison with the $\gamma$-ray
data can powerfully refine (or eliminate) the magnetosphere models.

\begin{deluxetable}{cccccccc}
\tablecaption{\label{tab3}Radio/Gamma-Ray Comparisons}
\tablewidth{0pt}
\tabletypesize{\scriptsize}
\tablehead{
\colhead{Pulsar}&\colhead{$\Psi$}&\colhead{$\zeta$}&
\colhead{$\Psi_R$}&\colhead{$\alpha_R$}&\colhead{$\beta_R$}&
\colhead{$\Delta_\gamma$}&\colhead{$\Delta_{OG}$}
}
\startdata
Crab        &124& 61& ?        & \nodata & \nodata & 0.39   & 0-0.35\\
Vela        &131& 64&$126.8\pm1$& 137 (43)\tablenotemark{a}& -6.5 & 0.42   & 0.35-0.4\\ 
J1930+1852  & 91&147& \nodata  & \nodata & \nodata &\nodata & 0 \\
J2229+6114  &103& 46&$\sim93$\tablenotemark{\dag}  & \nodata & \nodata &($\sim$0.4)\tablenotemark{b} & $\la 0.1$\\
B1706$-$44  &164& 53&$162\pm10$& \nodata & \nodata & 0.25   & 0-0.3\\	
J2021+3651  & 45& 79& \nodata  & \nodata & \nodata &($\sim$0.5)\tablenotemark{c} & 0-0.45\\
J0205+6449  & 90& 92& \nodata  & \nodata & \nodata &\nodata & 0-0.55 \\
J0537$-$6910&131& 93& \nodata  & \nodata & \nodata &\nodata & 0-0.55 \\
B0540$-$69  &144& 93& \nodata  & \nodata & \nodata &\nodata & 0-0.55 \\
J1124$-$5916& 16&105& \nodata  & \nodata & \nodata &\nodata & 0-0.45 \\
B1800$-$21  & 44& 90& \nodata  & \nodata & $\sim -6$\tablenotemark{d} &\nodata & $\sim$0.5 \\  
J1833$-$1034& 45& 85& \nodata  & \nodata & \nodata &\nodata &  0-0.55 \\
\enddata

\tablenotetext{\dag}{from the polarization structure of the radio nebula \citep{kot06}.}
\tablenotetext{a}{\citet{joh05}}
\tablenotetext{b}{\citet{tho02}}
\tablenotetext{c}{\citet{mcl03}}
\tablenotetext{d}{\citet{smi06}}
\end{deluxetable}

\section{Conclusion}

In conclusion, we have applied the torus fitting technique to more PWN observations in the
\emph{Chandra} data archive and characterized the uncertainties in the fits. This study
provides a better understanding of the systematic errors,
giving quantitative estimates of the measurement uncertainties. We argue that
these robust position angle $\Psi$ and inclination $\zeta$ values are particularly
useful for comparison with the radio and high energy pulse data. If new observations 
can fill in more measurements from these energy bands in Table 3, we should be able to make
substantial progress in understanding the emission zones and viewing geometries
of young pulsars.

\acknowledgments

This work was supported by NASA grant NAG5-13344 and by Chandra
grant AR6-7003 issued by the Chandra X-Ray Center, which is
operated by the Smithsonian Astrophysical Observatory for and
on behalf of the National Aeronautics Space Administration under
contract NAS8-03060.

{\it Facilities:} \facility{CXO (ACIS, HRC)}

\end{document}